\documentclass[11pt A4, twocolumn]{article}
\usepackage{amsfonts}
\usepackage{amssymb}
\usepackage{amsthm}
\usepackage{graphicx}
\usepackage{times}

\usepackage{setspace}
 \fontsize{11}{1}
\usepackage[left=2.5cm,top=2.5cm,right=2.5cm,bottom=2.5cm, nohead]{geometry}

\begin{document}
 \twocolumn[
\begin{@twocolumnfalse}
\begin{center}

\LARGE{Large-Area, Low-Noise, High Speed, Photodiode-Based\\
Fluorescence Detectors with Fast Overdrive Recovery} \normalsize

S. Bickman, D. DeMille

\textit{Yale University, Physics Department, PO Box 208120, SPL
23, New Haven, CT, 06520}

\bigskip

\end{center}
\begin{abstract}
Two large-area, low noise, high speed fluorescence detectors have
been built.  One detector consists of a photodiode with an area of
28 mm x 28 mm and a low noise transimpedance amplifier.  This
detector has a input light-equivalent spectral noise density of
less than 3 pW/$\sqrt{Hz}$, can recover from a large scattered
light pulse within 10 $\mu$s, and has a bandwidth of at least 900
kHz.  The second detector consists of a 16 mm diameter avalanche
photodiode and a low-noise transimpedance amplifier.  This
detector has an input light-equivalent spectral noise density of
0.08 pW/$\sqrt{Hz}$, also can recover from a large scattered light
pulse within 10 $\mu$s, and has a bandwidth of 1 MHz.
\end{abstract}

\end{@twocolumnfalse}
]
\section{Introduction} Two large-area, high speed, photodiode-based
fluorescence detectors have been built to detect fluorescence from
molecules excited via a pulsed laser system. Both detectors have a
large area to allow for collection of light from a large solid
angle and are fast to allow for good time resolution of the
fluorescence. Since the high intensity of pulsed laser systems
inevitably scatters a significant number of photons into the
detector, both detectors are designed to quickly recover from a
large pulse of light that would otherwise saturate the detector
for the duration of the fluorescence signal.

While photomultiplier tubes (PMTs) can also be used to measure
fluorescence after excitation with a pulsed laser, photodiodes and
avalanche photodiodes (APDs) have several advantages. Photodiodes
typically have a much higher quantum efficiency (q.e.) than PMTs.
For example, at 550 nm (where this detector is used) PMTs have
q.e.$\approx$20\% while Silicon PIN photodiodes have
q.e.$\approx$85\% and the APD used in this detector has
q.e.$\approx$80\%. Furthermore, photodiodes have a more linear
response to light intensity than PMTs and can withstand higher
sustained fluxes.

One of these detectors will be used in an experiment that intends
to improve the sensitivity to an electron electric dipole moment
(EDM)\cite{2004}.  The experiment detects small energy shifts in
an excited state of PbO that would result from a non-zero EDM. The
energy shifts are measured with quantum beat spectroscopy
\cite{Haroche}, which in this case appears as a sinusoidal
modulation at 200-500 kHz superimposed on an exponential decay due
to the spontaneous emission of an excited state of
PbO\cite{firstPbO}. In order to maximize the solid angle of
detection, large area photodiodes or APDs are used.  The Hamamatsu
S3584-08 silicon PIN photodiode has an area of 28 mm x 28 mm and
the Advanced Photonix 630-70-73-500 APD has a diameter of 16 mm.

While PMTs have an intrinsic nearly noise-free gain stage, PIN
photodiodes do not and the gain in APDs is smaller than the
desired gain. Thus, low noise transimpedance preamplifiers are
necessary for the photodiode-based detectors. There were three
requirements for the design of the preamplifiers in our
experiment. First, the intrinsic noise of the amplifiers must be
less than the anticipated shot noise, so that the amplifier noise
will not significantly contribute to the overall noise on the
detected signal.  Second, since the photodiode or APD is exposed
to a large pulse of scattered light from the excitation laser, the
preamplifier must be able to recover quickly from such a pulse. We
specifically require recovery in $\lesssim$10 $\mu$s, since the
excited state has a lifetime of 50 $\mu$s under the current
conditions. Finally, to allow for unattenuated detection of
quantum beats at $\sim$500 kHz, the preamplifier must have a
bandwidth of $\gtrsim$1 MHz.

\section{Low noise amplification}
\begin{figure}
\center
\includegraphics{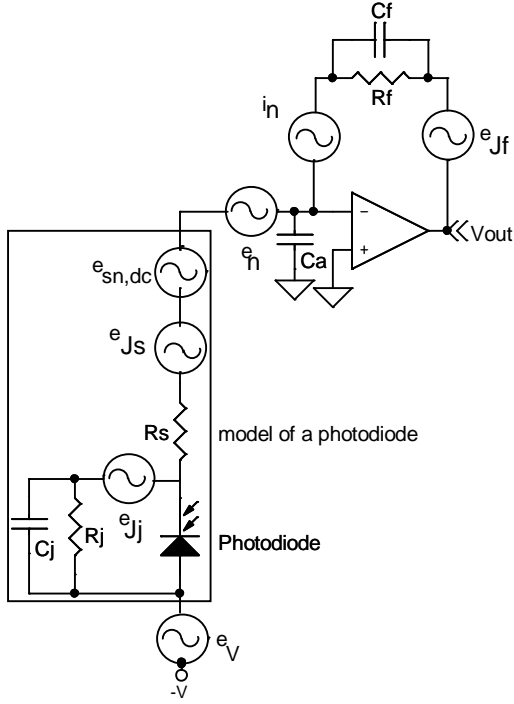}
\caption{Noise model of a transimpedance amplifier. Here, a real
photodiode is modelled as an ideal photodiode in parallel with a
junction capacitance $C_j$, a shunt resistance $R_j$ and a series
resistor $R_s$.}
 \label{fig:2}
 \end{figure}
 The noise of a transimpedance amplifier
can be modelled as shown in figure \ref{fig:2}\cite{PD}.  All
noise components will be calculated at the output of the
transimpedance amplifier.  A real photodiode can be modelled as an
ideal photodiode in parallel with a capacitor $C_{j}$ and a shunt
resistor $R_{j}$, plus resistance $R_s$ in series with the other
components.  $C_f$ and $R_f$ are the feedback capacitance and
resistance and $C_a$ is the amplifier input capacitance.

All of the resistances in this model have intrinsic Johnson noise.
The voltage noise spectral density $e_J$ across a resistance $R$
is given by $e_{J}=\sqrt{4 k_{B} T R}$. At the output of the
amplifier, the Johnson noise of $R_j$ contributes
\begin{equation}
e_{out, Jj}=\frac{\frac{R_f}{1+s C_f R_f}}{R_j+R_s}e_{Jj},
\end{equation}
 where $C_i=C_f+C_a$ and $s=i \omega$.
At the output of the amplifier, the Johnson noise from $R_s$ gives
\begin{equation}
e_{out, Js}=\left( 1+\frac{\frac{R_f}{1+s C_f
R_f}}{R_s+\frac{R_j}{1+s C_i R_j}} \right) e_{Js},
\end{equation}
and the Johnson noise from $R_f$ gives
\begin{equation}
e_{out, Jf}=\frac{e_{Jf}}{1+C_f R_f s}.
\end{equation}

In addition to Johnson noise, we must consider the voltage and
current noise at the input of the amplifier.  At the output of the
amplifier, the input voltage noise of the amplifier $e_n$
contributes
\begin{equation}
e_{out, n}=(1+\frac{\frac{R_f}{1+s C_f R_f}}{R_s+\frac{R_j}{1+s
C_i R_j}})e_n.
\end{equation}
The input current noise $i_n$, at the output of the amplifier,
gives
\begin{equation}
e_{out,i}=\frac{R_f i_n}{1+C_f R_f s}.
\end{equation}

In this application, it is necessary to keep the total electronic
noise less than the shot noise.  The expected signal size is
$\dot{N}=2 \times 10^{8}$ photoelectrons, in an exponential decay
with a time constant $\tau$ of $50 \mu s$.  For a signal current
$I_{sig}=\frac{\dot{N} e}{\tau} exp(-t/\tau)$, the current noise
spectral density at the input of the detector is
\begin{equation}
i_{sn}=\sqrt{2 e I_{sig}}\sqrt{F}G.
\end{equation}
Here $G$ is the intrinsic gain of the detector; $G$=1 for a PIN
photodiode, while $G$=200 for the APD used here.  Also, $F$ is an
additional noise factor associated with the gain process.  For the
PIN photodiode, $F$=1, while for an APD $F\geq 2$ \cite{Yariv}. In
the case of the APD used here, this additional noise was measured
as indistinguishable from F=2. At the output of the amplifier,
this becomes
\begin{equation}
\delta V_{sn}=R_f \frac{\sqrt{2 e I_{sig}}}{1+C_f R_f s}\sqrt{F}G.
\end{equation}
\subsection{PIN Photodiode}
 \begin{figure}
\center
\includegraphics{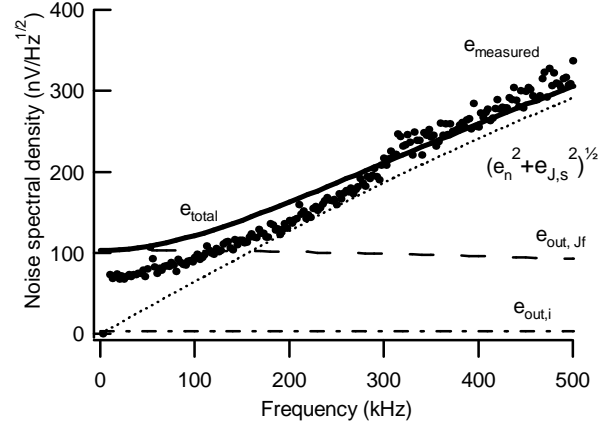}
 \caption{Graph of measured and predicted noise
referred to the output of the preamplifier for the PIN photodiode.
 For comparison, the predicted shot noise in our signals is 420 nV/$\sqrt{Hz}$.}
 \label{fig:99}
 \end{figure}
In the photodiode transimpedance amplifier, the feedback resistor
was chosen to be $R_f=$600 k$\Omega$. In this case, the shot noise
on the signal (at low frequencies) is $\delta V_{sig,sn}$=0.3
$\mu$V $exp(-t/2 \tau)$ (also at low frequencies) which is larger
than the Johnson noise of the feedback resistor for the first 100
$\mu$s of the decay since $e_{out, Jf}$=0.1 $\mu$V when $\omega
\ll \frac{1}{R_f C_f}$. Having chosen $R_f$=600 k$\Omega$, it is
possible to make some estimates of which noise terms are
important.  The Hamamatsu S3584-08 photodiode used to detect the
signals has a capacitance $C_j \approx 200 $ pF when reverse
biased and resistances $R_j \sim$ 5 G$\Omega$ and $R_s \sim 5
\Omega$. $C_f$ is approximately 0.3 pF in order to maintain the
desired bandwidth. With these values, the significant noise
sources are $e_{Js}$, $e_{Jf}$, and $e_n$, which all contribute on
the order of $10^{-7}$ V/$\sqrt{Hz}$ at the output. At high
frequencies, $e_{out,n} \approx e_n \frac{C_i}{C_f}$, so it is
necessary to keep $e_n \ll 1$ nV/$\sqrt{Hz}$. $e_{Jj}$ and $i_n$
contribute very little to the noise (see figure \ref{fig:99}).
\begin{figure}[h!]
\center
\includegraphics{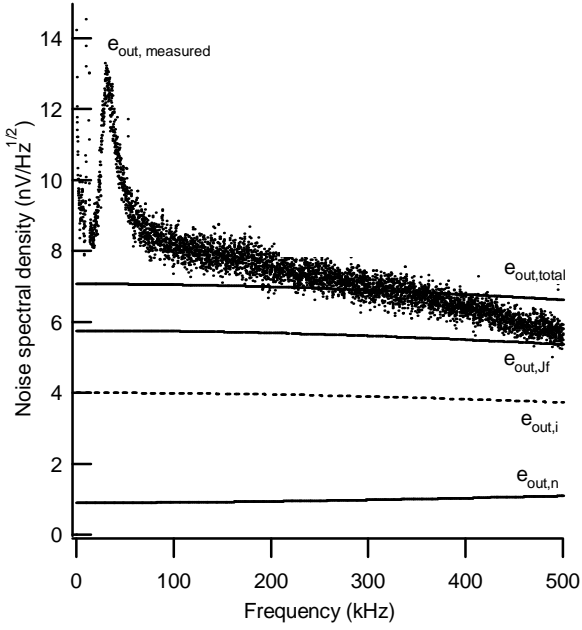}
 \caption{Graph of measured and
predicted noise referred to the output of the preamplifier for the
APD.
 For comparison, the predicted shot noise on our signal is 70 nV/$\sqrt{Hz}$.  The
 low frequency noise spikes appear to be a result of poor shielding on the
 APD.}
 \label{fig:100}
 \end{figure}
In order to control the input voltage and current noise of the
amplifier at the desired level, it is necessary to use a low noise
JFET as the the input stage of the amplifier.  Most low-noise
operational amplifiers have an input voltage spectral noise
 which is too high for this application.  Each
of the IF9030 JFETs used here has $e_n$= 0.5 nV/$\sqrt{Hz}$. These
JFETs do not have enough gain to achieve the desired
transimpedance gain of $\frac{V_{out}}{I_{in}}=6 \times 10^{5}$
$\Omega$, so this stage is followed by an operational
amplifier\cite{Stubbs}. The JFET input stage provides a low-noise
amplification stage before the noisier second stage op-amp.  The
gain of the JFET amplifier is sufficiently large that the input
noise of the op-amp is negligible.

The input voltage noise $e_n$ can be further reduced by putting
$m$ JFETs in parallel resulting in a combined noise of
$e_n$=$e_{JFET}/\sqrt{m}$. In principle, many JFETs can be used to
improve the noise of the amplifier, but in practice there are two
limitations.  The ultra-low noise JFETs used here have a
significant gate-source capacitance $C_{JFET}$. Since the noise at
the output of the amplifier at high frequencies scales as $e_n C_a
\propto \frac{1}{\sqrt{m}}(C_{j}+m C_{JFET})$, the number of
parallel JFETs has an optimum value $m\approx C_j/C_{JFET}$
\cite{HH}. Also, since $e_{J,s}$ is of the same order as $e_n$,
there is little improvement to the noise once $e_n \lesssim
e_{J,s}$.

Requirements for the second stage of amplification were also
stringent.  At high frequencies the closed-loop voltage gain of
the entire preamplifier is $G_{v,cl} \approx \frac{C_i}{C_f} \sim
1000$. However, the voltage gain of the JFET front-end is only
$G_{1} \sim 60$, determined by the JFET transconductance of
$g_m=0.02$ S and the drain resistance of $3$ k$\Omega$.  The
Miller effect significantly reduces the gain of the JFET stage at
high frequencies, so a cascode configuration was used to reduce
this effect\cite{JW}. Even with the cascode, the second stage
amplifier must have a high gain and wide bandwidth so that the
open loop gain is large enough.  We use the Analog Devices AD797,
which which has a gain bandwidth product GBW of 110 MHz, and has
$e_{n2}=0.9$ nV/$\sqrt{Hz}$.  In addition to the high GBW of the
AD797, this op-amp was chosen for its fast recovery from
overdrive, which allows it to quickly recovery from the scattered
light pulse.

\subsection{APD}
 The noise design of the transimpedance amplifier for the APD is
 much less stringent since the bias voltage is chosen so
 the APD gain G$\approx$200.  The shot noise at the output of the APD
 is thus $\sqrt{2} G$ larger than the noise in the PIN photodiode.  We choose $R_f$=2 k$\Omega$
  in the transimpedance amplifier for the APD, resulting in shot noise at
  the output of the amplifier $\delta V_{sn}=100 nV/\sqrt{Hz}$. Having chosen $R_f$ and
  knowing that $C_j$=140 pF for this APD,
 it is possible to make some estimates of which noise terms are
 significant.  We find that $R_j$ and $R_s$ have insignificant
 noise contributions.  Since the noise requirements are much less
 strict for this amplifier, no front-end JFET was used.  Instead the
 transimpedance amplifier is made from a
 single op-amp, the AD797, which has again been chosen for its
 high GBW and fast overdrive recovery time.  The AD797 has
 $e_n=0.9$ nV/$\sqrt{Hz}$ and $i_n=2.0$ pA/$\sqrt{Hz}$.  At the
 output of the transimpedance amplifier, $e_{out,Jf}\sim$ 5.8
 nV/$\sqrt{Hz}$.  The voltage noise is almost insignificant since $e_{out, n}
 \sim$ 1.3 nV/$\sqrt{Hz}$.  The current noise is the largest expected noise
 contribution at $e_{out, i}\sim$ 3.7 nV/$\sqrt{Hz}$.  The various
 noise contributions along with the measured noise are shown in
 figure \ref{fig:100}.

There are two other noise sources that are significant with the
APD detector, but not the photodiode detector.  The first is the
shot noise in the dark current of the APD $e_{sn,dc}$.  In this
case, this shot noise is negligible compared to the shot noise on
the expected signal. The second is the noise on the bias voltage
for the APD. The photodiode bias voltage was supplied by
batteries, but the necessary high bias voltage for the APD makes
it difficult to supply with batteries. A high voltage power supply
was used and filtered to provide acceptable noise characteristics
for the APD detector.

\section{Recovery time from scattered light pulse} Many experiments
observe the fluorescence from atoms or molecules that are excited
with a pulsed laser. These high intensity beams scatter photons
into the detector causing a large, but temporally short burst of
photoelectrons.  In many cases, this scattered light pulse is
several orders of magnitude larger than the expected signal size,
and can easily saturate the detector system.  In our case, the
transimpedance amplifier must be able to recover from this
scattered light pulse, which is injected in a $\sim$5 ns duration,
quickly in order to observe the spontaneous fluorescence.  We
define the recovery time as the time that it takes the amplifier
to ring down to 1/e of the saturated value.

\begin{figure}
\center
\includegraphics{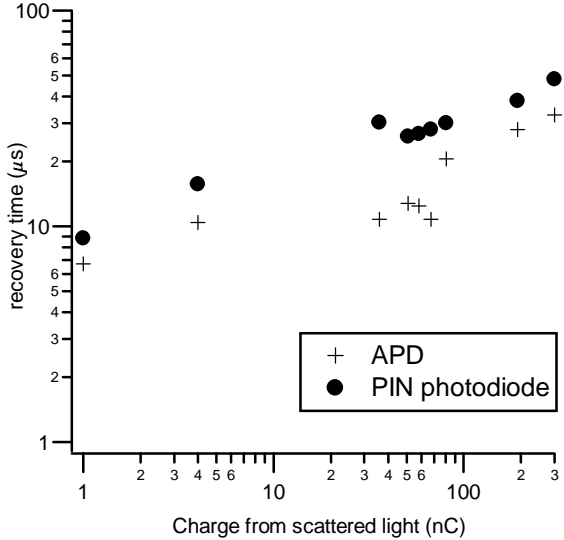}
\caption{Graph of the recovery time of the detectors
 vs the size of the scattered light pulse.} \label{fig:80}
\end{figure}

The recovery time of the amplifier depends on the size of the
scattered light pulse.  For this experiment, it was necessary to
choose optical filters that reduced the signal size, but also
reduced the scattered light.  The two filtering options we
considered allowed for a scattered light pulse that injected a
charge of either 220 nC or 1.5 nC at the expense of a factor of 3
smaller desirable signal in the latter case.  The photodiode and
APD both took more than 50 $\mu$s to recover from the 220 nC pulse
and $\approx 10 \mu$s to recover from the 1.5 nC pulse.  The
recovery time as a function of injected charge from scattered
light is shown in figure \ref{fig:80} for both detectors.

Three changes were made to the design of the transimpedance
amplifiers to improve the recovery time.  First, clamping diodes
were placed between the output of the photodiode or APD and ground
to shunt large input light signals to ground.  Small signal and
Schottky diodes were used and both kinds were chosen for their low
capacitance so as not to significantly contribute to $C_i$.  The
Schottky diodes were chosen for their low forward voltage drop,
and the small signal diodes were chosen for their higher current
ratings.  These diodes are placed in opposite directions to damp
additional electronic ringing due to the scattered light pulse.

The second and third changes were only applied to the photodiode
amplifier.  Here, diodes were placed in the feedback loop to allow
for a low resistance path when the amplifier was saturated.  These
diodes were not necessary in the APD amplifier because $R_f$ is
much smaller.  The cascode in the JFET stage improved the recovery
time by increasing the bandwidth of this stage of the amplifier.

Preliminary circuit designs used two additional methods of
improving the recovery time, but these methods were not successful
enough to be used in the final design.  The first method was to
add a power booster in the feedback loop for the photodiode
detector.  This power booster was able to source more current
through the diodes in the feedback loop of this detector in an
attempt to shorten the recovery time.  In this application, it was
difficult to implement the power booster as it made the feedback
traces larger, and $C_f$ difficult to minimize.  The second method
was to add an additional photodiode at the input of the
transimpedance amplifier with opposite orientation to the
detecting photodiode.  A laser diode shone into the additional
diode providing some charge cancellation during the scattered
light pulse.  However, the improvement in recovery time was found
to be negligible.
 \begin{figure}
\center
\includegraphics{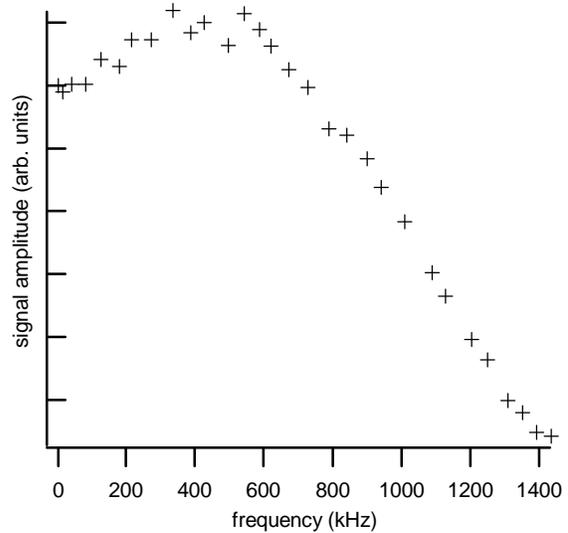}
\caption{Measured frequency response of the PIN photodiode
transimpedance amplifier.} \label{fig:12}
\end{figure}
\begin{figure*}[p]
\center
\includegraphics{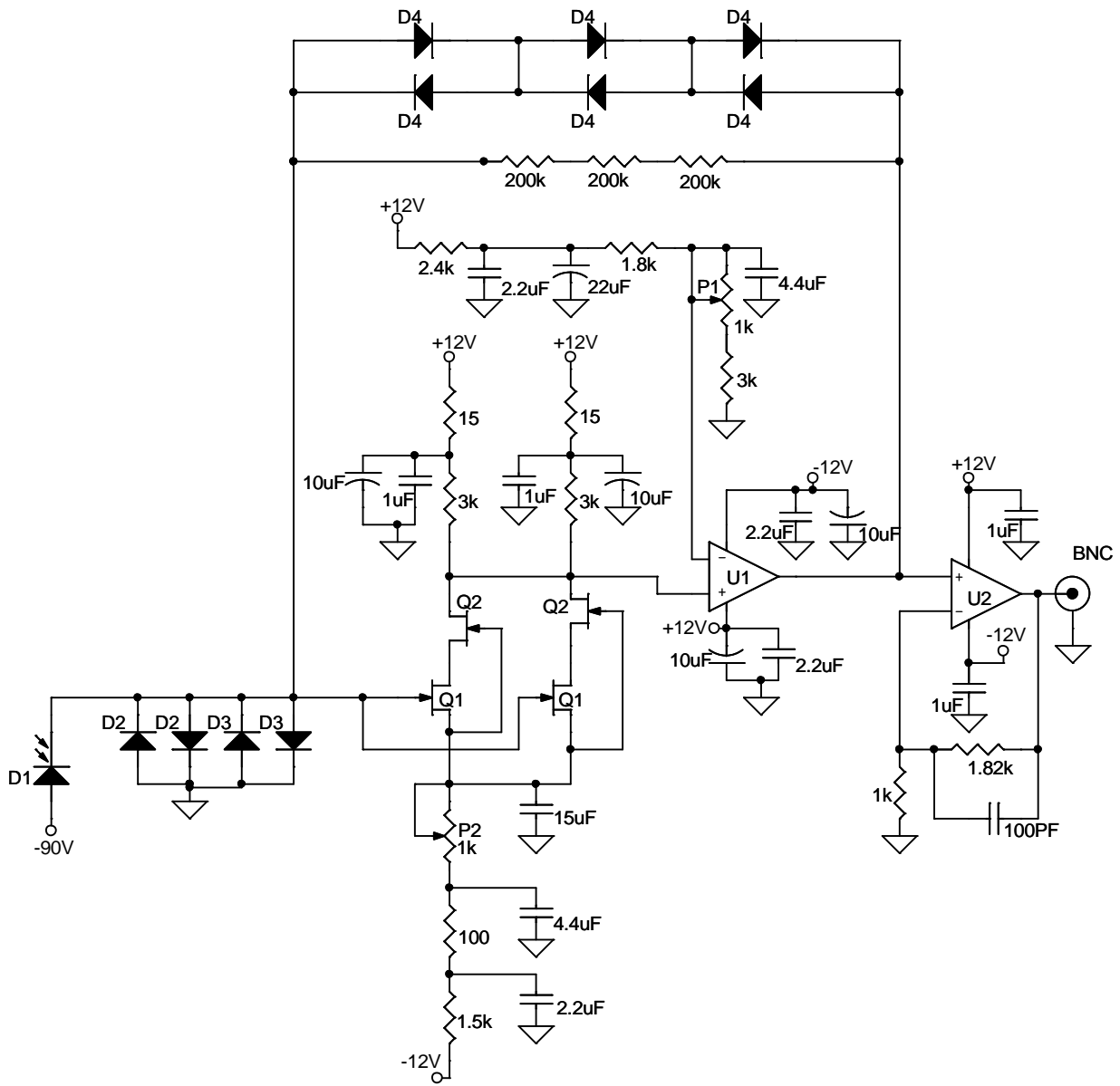}
 \caption{Circuit diagram for the PIN photodiode
detector.
 D1 is the large-area Hamamatsu S3584-08 photodiode.  D2 are BAS70-04 Schottky diodes chosen for their
 small capacitance and low forward voltage.  D3 are small signal HSMP-3822-BLK
 diodes, which allow for more current conductance than the Schottky diodes.
 D4 are also HSMP-3822-BLK diodes, chosen for their low capacitance.  When the preamplifier is
 saturated, these diodes provide a low resistance path in the feedback loop.  Since the JFETs
 Q1 and Q2 drift with temperature, P1 and P2 are
used to finely adjust the voltages
 at the inputs of U1 so that the two inputs are at the same DC voltage.  Q1 is chosen for its low voltage noise, and is an IF9030 from
InterFET.  Q2 is used in a cascode
 configuration to increase the speed of the JFETs and is a 2N4856A from InterFET.  U1 is an AD797, which was chosen for its
 high speed, low noise, and fast overdrive recovery.  U2 is an AD829, which is a cable driver chosen for its fast overdrive
 recovery.}
 \label{fig:9}
 \end{figure*}

  \begin{figure*}[p]
\center
\includegraphics{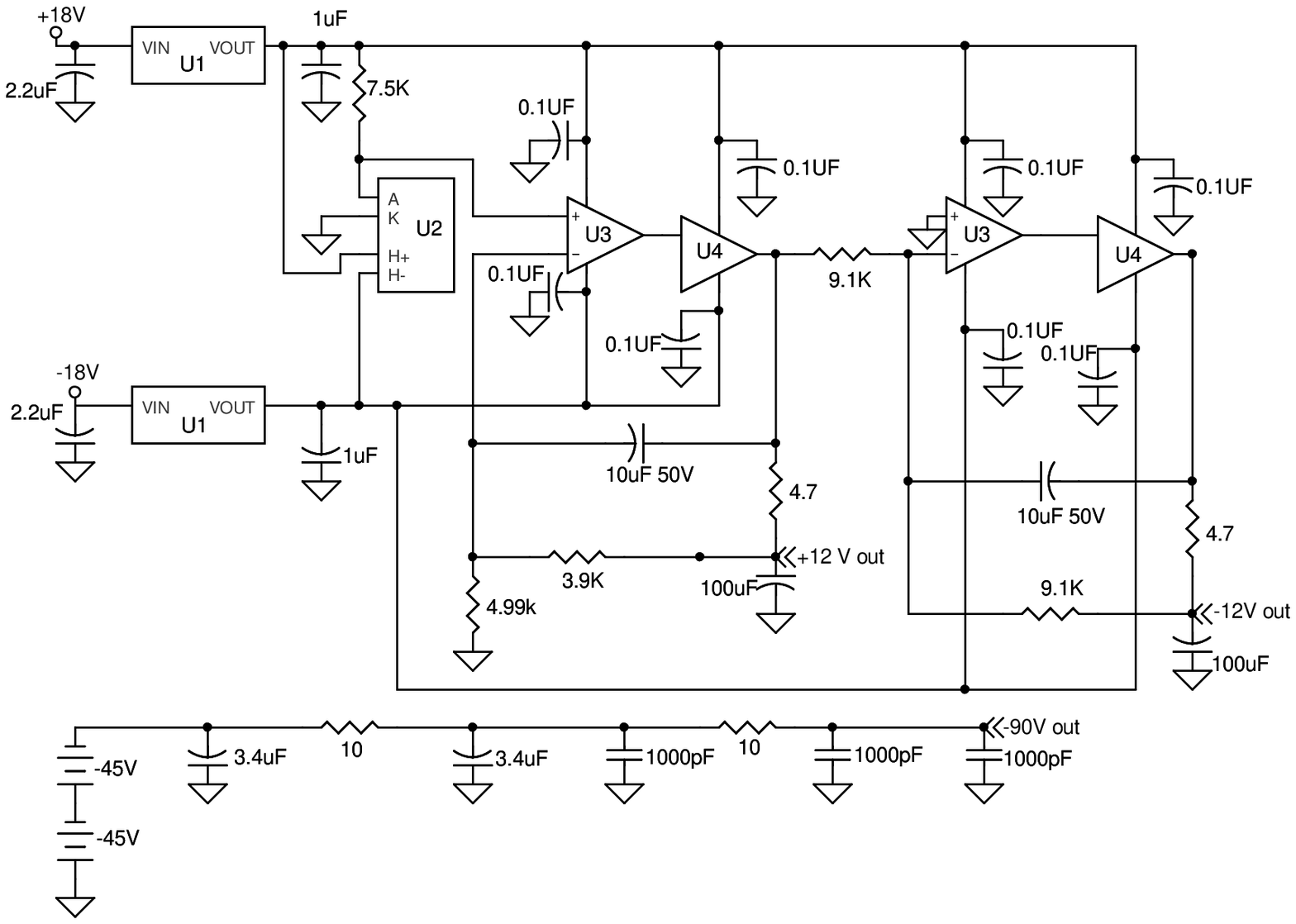}
 \caption{Power supplies for preamplifier.  The $\pm 18V$ power supplies powering these components are powered
by a PowerOne AC/DC converter.  This circuit can supply 200 mA
continuously, and up to 500 mA for short times.  The noise on
these power supplies is less than 2 nV/$\sqrt{Hz}$ for all
frequencies between 10 Hz and 1 MHz.  U1 are 7815 and 7915 voltage
regulators.  U2 is a LM399H voltage reference, which was chosen
for its stability. U3 are low-noise OP77 op-amps, and U4 is a
BUF634 power-boosting amplifier.  The two -45 V power supplies are
batteries.}
 \label{fig:10}
 \end{figure*}
\section{Frequency Response}

In order to achieve the desired bandwidth of $\sim$1 MHz with a
feedback resistor of 600 k$\Omega$ in the PIN photodiode
transimpedance amplifier, it is necessary to keep the feedback
capacitance less than 0.26 pF. Since the diodes used to decrease
the scattered light recovery time have a combined capacitance of
0.2 pF, it is necessary to keep all other possible capacitances
extremely low. For this reason, three 200 k$\Omega$ resistors were
used in series to create the feedback resistor. Additionally, the
circuit board was designed to keep the traces relating to the
components in the feedback loop as short as possible by placing
feedback components on the opposite side of the circuit board from
the amplifying components. Ground planes were placed on all
available surfaces on the circuit board and two inner layers of
the circuit board were also grounded.  The 3dB bandwidth of the
circuit was measured as $\approx$900 kHz, implying $C_f\approx$
0.3 pF (see figure \ref{fig:12}).

The smaller $R_f$ in the APD amplifier makes it much less
sensitive to frequency limitations and this amplifier could be
made to have a bandwidth of at least 10 MHz if necessary. Our
present APD detector has a bandwidth of 1 MHz.

\section{Final Design}

The final design incorporated all of the elements discussed above.
The circuit diagram for the PIN photodiode detector is shown in
figure \ref{fig:9}. The circuit boards were designed with great
care to minimize the path lengths at the input of the amplifier.
These short paths reduce capacitive effects and noise pickup from
other sources. Surface mount components were used wherever
possible to reduce lead size. All cables leading to or from the
circuit board are made of coaxial cable.  The power supplies for
the photodiode detector were also constructed to minimize noise.
The design of the power supplies is shown in figure \ref{fig:10}.

\begin{figure}
\center
\includegraphics{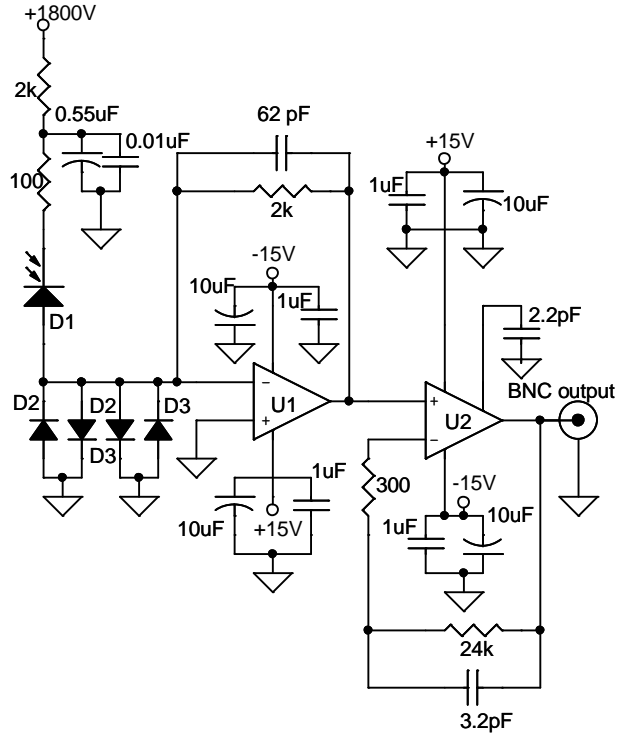}
 \caption{Circuit Diagram for the APD detector. D1 is
the APD, and is the
 630-70-73-500 from Advanced Photonix, chosen for its large area.
 D2 is a BAS70-04 Schottky diode, chosen for its low forward voltage
  and low capacitance.  D3 is a HSMP-3822-BLK diode chosen for low capacitance.
  U1 is an AD797, chosen for low-noise, high speed and fast overdrive recovery.
  U2 is an AD829, a high speed cable driver with fast
overdrive recovery.}
 \label{fig:999}
 \end{figure}

The final design for the APD detector is shown in figure
\ref{fig:999}.  The power supply for the bias voltage was filtered
to remove high frequency noise components.

\section{Summary}
The large-area (7.8 cm$^2$) PIN photodiode detector described here
has an input light-equivalent noise of less than 3 pW/$\sqrt{Hz}$
at a wavelength of 550 nm, can recover in 10 $\mu$s from large
scattered light pulses that rapidly inject 1.5 nC of charge, and
has a bandwidth of more than 900 kHz. The noise was minimized by
using a large feedback resistance to minimize Johnson noise, JFETs
as the first stage of the amplifier to $e_n$, and low noise power
supplies.  The recovery time from the scattered light was reduced
by using clamping diodes on the input and in the feedback, and
with a cascode configuration on the JFET front end. The bandwidth
was achieved by minimizing all capacitances.

\begin{figure}
\center
\includegraphics{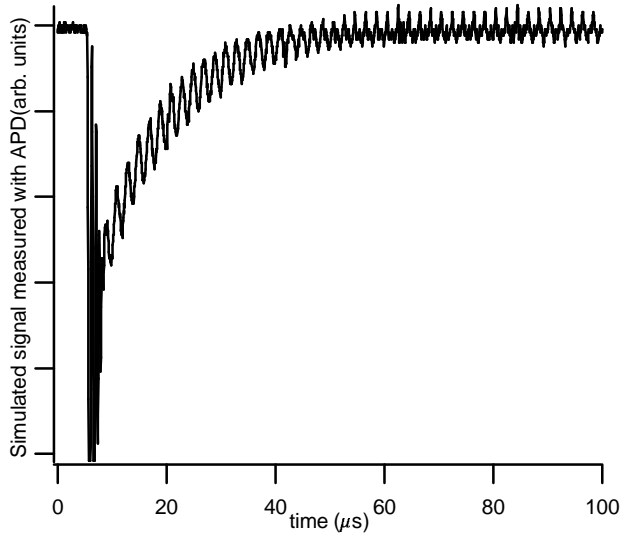}
 \caption{Simulated quantum beats at 500 kHz measured
with the APD detector and scattered light equivalent to 1.5 nC of
injected charge.}
 \label{fig:11}
 \end{figure}

The final design for the large-area (2.0 cm$^2$) APD detector has
an input light-equivalent noise of 0.08 pW/$\sqrt{Hz}$, can
recovery quickly from large scattered light pulses and easily has
a bandwidth of 1MHz.  Quantum beats were observed with this APD
detector; however, the signals in the current experimental
configuration are much smaller than ultimately anticipated.  In
figure \ref{fig:11}, simulated signals at the anticipated level of
0.6 $\mu$A, with quantum beats at 500 kHz, are shown as measured
with the APD. These signals were simulated with an LED driven by
an arbitrary waveform generator programmed to simulate the beat
signal. The trace in figure \ref{fig:11} includes scattered light
equivalent to 1.5 nC of charge injection.

\section{Acknowledgements}

 We are grateful for the support of NSF
Grant No. PHY0244927, and the David and Lucile Packard Foundation.
 We thank David Kawall and Valmiki Prasad for
helpful discussions.

\bibliographystyle{srt}

\begin{thebibliography}{99}
    \bibitem{2004}D. Kawall, F. Bay, S. Bickman, Y. Jiang, and D. DeMille
Phys. Rev. Lett. \textbf{92}, 133007 (2004)

 \bibitem{Haroche} S. Haroche in \underline{High-Resolution Laser Spectroscopy} edited by K. Shimoda (Springer-Verlag,
     Berlin, 1976), Chap. 7, pp 253-313.

     \bibitem{firstPbO}D. DeMille, F. Bay, S. Bickman, D. Kawall, D. Krause, Jr., S. E. Maxwell, and
     L. R. Hunter, Phys. Rev. A \textbf{61}, 052507 (2000).

     \bibitem{PD} J. Graeme, \underline{Photodiode Amplifiers: Op-Amp Solutions}.  McGraw-Hill, New
York 1996.


  \bibitem{Yariv} Ammon Yariv. \textit{Optical Electronics}, Third
  Edition, CBS College Publishing, USA (1985).

  \bibitem{Stubbs} D.Yvon, A. Cummings, W. Stockwell, P. Barnes,
  C. Stanton, B. Sadoulet, T. Schutt, C. Stubbs, \textit{Nucl.
  Instr. and Meth. in Phys. Res. A} \textbf{368} (1996) 778-788.

  \bibitem{HH} Paul Horowitz, and Winfield Hill. \textit{The Art of
    Electronics}, Second Edition, Cambridge University Press, USA
    (1998).

    \bibitem{JW} Steven R. Jefferts, and F.L. Walls. \textit{Rev.
    Sci. Instru.} \textbf{60} (1989) 1194-1196.

\end{thebibliography}

\end{document}